\documentclass[acmsmall,screen]{acmart}

\AtBeginDocument{%
  }

\setcopyright{acmlicensed}
\copyrightyear{2018}
\acmYear{2024}
\acmDOI{XXXXXXX.XXXXXXX}

\acmJournal{JACM}
\acmVolume{37}
\acmNumber{4}
\acmArticle{111}
\acmMonth{8}

\usepackage{enumitem}
\usepackage{url}
\usepackage{multirow}
\usepackage{geometry}
\usepackage{booktabs} 
\usepackage{geometry}
\usepackage{amsmath}
\usepackage{graphicx}
\usepackage{cleveref}
\usepackage{subcaption}
\usepackage{listings}
\usepackage{threeparttable}
\usepackage{xcolor}
\usepackage{verbatim}
\usepackage{bbding}
\usepackage{pifont}
\usepackage{amsfonts}
\usepackage{wasysym}
\usepackage{utfsym}
\usepackage{colortbl}
\usepackage[many]{tcolorbox}
\usepackage{fontawesome5}
\usepackage[ruled,linesnumbered]{algorithm2e}
\usepackage{algpseudocode}

\crefname{algocf}{algorithm}{algorithms}
\Crefname{algocf}{Algorithm}{Algorithms}

\newtcolorbox{boxK}{
    top=2pt,
    bottom=2pt,
    left=2pt,
    right=2pt,
    boxrule = 0pt,
    toprule = 0pt, 
}

\definecolor{highlight}{HTML}{C0C0C0}

\begin{document}

\title[Enhancing LLMs in Long Code Translation through Instrumentation and Program State Alignment]{Enhancing Large Language Models in Long Code Translation through Instrumentation and Program State Alignment}

\author{Xin-Ye Li}
\email{lixy@lamda.nju.edu.cn}

\author{Ya-Li Du}
\email{duyl@lamda.nju.edu.cn}

 \author{Ming Li}
\email{lim@lamda.nju.edu.cn}
\affiliation{
  \institution{National Key Laboratory for Novel Software Technology, Nanjing University}
  \country{China}
 }

\affiliation{
  \institution{School of Artificial Intelligence, Nanjing University}
  \country{China}
 }



\begin{abstract}
Code translation aims to transform code from one programming language to another while ensuring functional equivalence. It has board applications in cross-platform development and software migration, such as transitioning C++-written systems to Rust. Recently, code translation has been greatly enhanced by the profound advancements in Large Language Models (LLMs). 
However, despite these improvements, existing LLM-based approaches struggle to infer a program's functionality from its appearance, leading to failure in preserving the program semantics during translation. This challenge becomes even more pronounced when dealing with longer and more complex code: current code LLMs face difficulties in handling lengthy programs and have limited capability in understanding intricate semantics. This limitation is particularly critical, as translating long code is more practical in real-world scenarios. To evaluate the capability of recent code LLMs on long code translation, we introduce \textit{LongTrans}, a large-scale execution-based code translation benchmark. \textit{LongTrans} consists of C++, Java, and Python programs, with lengths ranging from hundreds to thousands of tokens. 
Our empirical study on 12 series of LLMs reveals a dramatic performance decline as code length increases, regardless of model size. Even the most performant LLM, GPT-4o, achieves only 57.51\% computational accuracy, highlighting the pressing need for further research in long code translation.

In this paper, we argue that the essence of code translation is maintaining invariant functionality while transforming its appearance (e.g. syntax, keywords) from one programming language to another.
Although the translated code differs in appearance from the original, its program states should remain identical to the original from entry to exit throughout execution. To achieve this goal, we propose aligning program states at runtime beyond the final output to ensure functional consistency, ultimately producing an accurate translation. 
We propose an approach termed \textbf{P}rogram \textbf{S}tate \textbf{A}lignment augmented \textbf{T}ranslation (\textbf{PAST}), to augment code LLMs with instrumentation, a classical technique in dynamic program analysis, to capture and describe the program states and assist their alignment during translation. 
Specifically, it is the first attempt to utilize code LLM to insert instrumentation code into both the original and the translated code and trace their program states at runtime. 
We prompt the code LLM to identify and repair the error based on the output traces, mitigating the inconsistency and thereby improving the accuracy of the translation. Extensive experimental results demonstrate that our approach significantly improves the performance of code LLMs in long code translation, with the computational accuracy increasing from 57.51\% to 84.70\% for GPT-4o, increasing from 50.68\% to 69.97\% for Mistral-Large-2, and increasing from 52.45\% to 76.43\% for DeepSeek-Coder-V2. Moreover, the results also show that these substantial improvements are consistent across different code LLMs and translation datasets, such as C++ to Java. Ablation studies further confirm that these improvements are brought by incorporating instrumentation and program state alignment.
\end{abstract}

\ccsdesc[300]{Software and its engineering~Automatic programming}

\keywords{code translation, large language model}


\maketitle

\section{Introduction}

Code translation aims to transform programs from one language to another while preserving their original functionality. It effectively facilitates cross-language migration and enables organizations to modernize legacy systems for better performance, maintainability, and scalability~\cite{Liu2023, Krishna2021, Haugeland2021, NguyenNN15}. As software has evolved, various programming languages have been developed to address diverse needs, such as desktop, web, and mobile applications. This growing diversity has made it increasingly necessary to port software between languages to support expanding business platforms~\cite{Wu2010, Gholami2017, AllamanisBDS18}. To meet this demand, different source-to-source translators, or transpilers, have emerged over the past few decades, improving migration efficiency and enhancing interoperability across large companies using multiple languages~\cite{KaraivanovRV14, NguyenNN16a, transcoder, Roziere2022}. Automated code translation techniques, thus, remain essential in accelerating migration, reducing costs, and boosting development efficiency.

In recent years, learning-based approaches have achieved impressive performance on code-related tasks, such as code summarization~\cite{Feng2020, Guo2020, Ahmad2020, Chai2022}, code generation~\cite{Iyer2018, Clement2020, Lu2021}, bug localization~\cite{sgattention,semanticcodebert}, and clone detection~\cite{Wei2017, Zhang2019}. Additionally, learning-based methods also have been proposed to enhance the efficiency and effectiveness of code translation~\cite{Chen2018, transcoder, Roziere2022, transcoder3, du2024joint}, typically leveraging task-specific pre-training on large monolingual corpora. Although these approaches have led to notable improvements, their current performance remains insufficient for practical deployment. 
Recent advances in Large Language Models, pre-trained on billions of text and code tokens, present an alternative solution by eliminating the need for re-training or fine-tuning while demonstrating strong capability across various code-related tasks~\cite{Chen2021, Li2023, Liu2024, Yang2024, Fan2023, Geng2024}. Furthermore, recent studies have explored the performance of LLMs of code on code translation, showing promising results~\cite{yang2024exploring, Tang2023, Jiao2023}. 

However, existing LLM-based approaches are translating code by appearance instead of inferring the underlying functionality and transforming the appearance from one programming language to another. Thus these approaches face challenges in preserving the program semantics during translation. 
Moreover, existing code translation benchmarks are typically limited to short code, such as snippet-level or function-level code, restricting their applicability in practice. In contrast, translating long code -- at the program or repository level -- is more complex and more common in real-world scenarios~\cite{Pan2023}. LLMs of code struggle to capture complex functionality from the appearance of long code inputs, even with extended context windows. Additionally, they are more prone to errors when generating long code outputs. Addressing these challenges is critical given the dominance of long code in practical software development.

To assess the capability of recent LLMs of code on long code translation scenarios, we conduct an empirical study on 12 series of LLMs, including GPT-4o~\cite{gpt4},  Llama-3.1~\cite{touvron2023llama}, Mistral-Large-2~\cite{jiang2023mistral}, and DeepSeek-Coder~\cite{zhu2024deepseek}, across six translation tasks, i.e. C++ to Python, Python to C++, Java to C++, C++ to Java, Java to Python and Python to Java. First, we construct an execution-based code translation benchmark called \textit{LongTrans}, which consists of a large collection of programs in C++, Java, and Python. Based on \textit{LongTrans}, we evaluate the above models under the metric of Computational Accuracy (CA). The results reveal a significant performance drop as the length of input code increases, with even the most capable LLM, GPT-4o, achieving only 57.51\% computational accuracy, which is far from satisfactory. 
To better understand the causes of translation errors, we further analyze the failed samples in the translation results. Interestingly, whether due to a runtime error or a wrong answer, most failed samples tend to fail across all tests. 
This finding suggests that the translation errors produced by LLMs of code are relatively superficial, reinforcing our assumption that LLMs of code have difficulty inferring the program's underlying functionality from its appearance. 
Recent work~\cite{yang2024exploring} attempts to repair the translation errors solely based on the execution feedback, such as error messages. However, identifying and repairing these errors in long code, which can span hundreds and thousands of lines, remains significant for current LLM-based approaches, akin to finding a needle in a haystack.

To address this challenge, we propose that the core of code translation lies in preserving the program's functionality and runtime behavior while transforming its appearance from one programming language to another. Although the program's appearance changes during translation, the translated program's states should remain consistent with the original from entry to exit throughout execution. To achieve this consistency, we propose a new approach termed \textbf{P}rogram \textbf{S}tate \textbf{A}lignment augmented \textbf{T}ranslation (\textbf{PAST}) to enhance the LLMs of code in long code translation through instrumentation and program state alignment. Our approach improves code translation by leveraging three key stages: Instrumentation, Translation, and Program State Alignment. In the Instrumentation stage, we insert the instrumentation code into the original program to monitor its program states and track the execution flow. During the Translation stage, LLMs of code translate the original program while preserving the insert instrumentation. If the translated program fails to maintain consistency, we proceed to the Program State Alignment stage. In this final stage, we execute both the original and translated programs on public test inputs and address any detected misalignments through repair. We incorporate two repair strategies: direct repair and localize-and-re-translate, effectively enabling our approach to align the program states thus achieving correct translation.

We highlight our contributions in four key aspects:
\begin{enumerate}
\item We propose that the core of code translation is to maintain the original program's functionality while transforming the appearance from one programming language to another. This consistency is essential, as existing LLM-based methods struggle to infer the underlying functionality solely from code appearance.
\item We highlight the importance of long code translation and introduce \textit{LongTrans}, a large-scale execution-based code translation benchmark containing C++, Java, and Python programs of varying complexities and lengths. Moreover, we conducted a comprehensive empirical study on 12 series of LLMs and carefully analyzed the results. Experimental findings reveal a significant decline in performance among LLMs of code as input code length increases, regardless of model size or programming language.
\item We propose a novel approach named \textbf{PAST} that utilizes instrumentation to capture the program states and trace the  execution flow. By leveraging instrumentation, LLMs of code can effectively identify and address inconsistencies between the program states of both the original and the translated program, resulting in improved translation accuracy.
\item Experimental results demonstrate that our approach substantially enhances LLMs of code in long code translation, with the computational accuracy increasing from 57.51\% to 84.70\% for GPT-4o, increasing from 50.68\% to 69.97\% for Mistral-Large-2, and increasing from 52.45\% to 76.43\% for DeepSeek-Coder-V2. Comprehensive evaluation further confirms that our proposed approach's effectiveness is consistent across code LLMs and translation datasets.
\end{enumerate}

\section{Background and Related Work}

\subsection{Automatic Code Translation}

The automated code translation tool aims to construct a function $f(x)$, that maps source code $x$ to target code $y=f(x)$. Early approaches to code translation were primarily rule-based, applying techniques such as phrase-based statistical machine translation for program translation~\cite{KaraivanovRV14,NguyenNN15,NguyenNN13,NguyenNN16a,AllamanisBDS18,swt}, which leveraged the grammatical structures of programming languages for code migration.

In recent years, various deep learning techniques have been employed for program translation~\cite{LampleCDR18,babel,pt5,du2024joint}. Depending on whether the training data is monolingual or bilingual, these approaches can be categorized into supervised code translation~\cite{must,pscpt} and unsupervised/weakly supervised translation~\cite{transcoder,transcoder2,transcoder3}. The latter focuses on using runtime testing with test cases, which is more valuable in practical development.
Additionally, with the rise of large language models (LLMs)~\cite{yang2024exploring}, code translation based on direct inference from LLMs has garnered widespread attention. Despite their remarkable achievements, these models still have clear limitations: (1) translating long code is challenging; (2) the time and space overhead for executing long code is large, leading to high evaluation costs; and (3) it is difficult to localize and repair defect when errors occur in execution accurately, especially for long code.

\subsection{Large Language Models for Code}
LLMs have demonstrated impressive performance on a variety of natural language processing (NLP) tasks~\cite{gpt4} and have also attracted broad research interest. These LLMs are pre-trained on trillions of tokens, which often include a significant portion of code tokens sourced from open-source code-hosting platforms and developer communities, such as GitHub\footnote{https://github.com/} and StackOverflow\footnote{https://stackoverflow.com/}. As a result, many LLMs exhibit non-trivial code understanding and generation capability even though they are not specifically trained for code.

Inspired by the remarkable success of LLMs in natural language processing, researchers have turned their attention to enhancing the coding capabilities of LLMs. They discovered that continual pre-training on code significantly improves LLMs' performance on code-related tasks. \mbox{\citet{Chen2021}} proposed Codex, a series of GPT-3 checkpoints pre-trained on 100B additional code tokens. Codex demonstrated a substantial improvement in code generation, solving 28.8\% of hand-crafted programming problems whereas GPT-3 solved 0\%. Moreover, existing works suggest that the performance of code generation can be further improved through self-consistency~\cite{Wang2022}, candidate ranking~\cite{Zhang2023, Inala2022, Lyu2024}, and other advanced prompting strategies. Besides code generation, LLMs also show great potential in other tasks, such as program repair~\cite{Jiang2023}, vulnerability detection~\cite{Steenhoek2023}, and code review~\cite{Lu2023}.

\section{Motivation}
\label{section:motivation}

This section presents the motivation behind our proposed approach. Our motivation is derived from a comprehensive empirical study conducted on our \textit{LongTrans} benchmark. In this study, we assess the performance of recent code LLMs with varying parameter sizes across six translation datasets involving the mutual translation between C++, Java, and Python. The evaluation of translated programs is based on execution results, with the details described in \Cref{subsection:evaluation_metrics}. Therefore, evaluation metrics are execution-based, focusing on the compilation success rate and pass rate. 

\begin{figure}[ht]
    \centering
    \includegraphics[width=\textwidth]{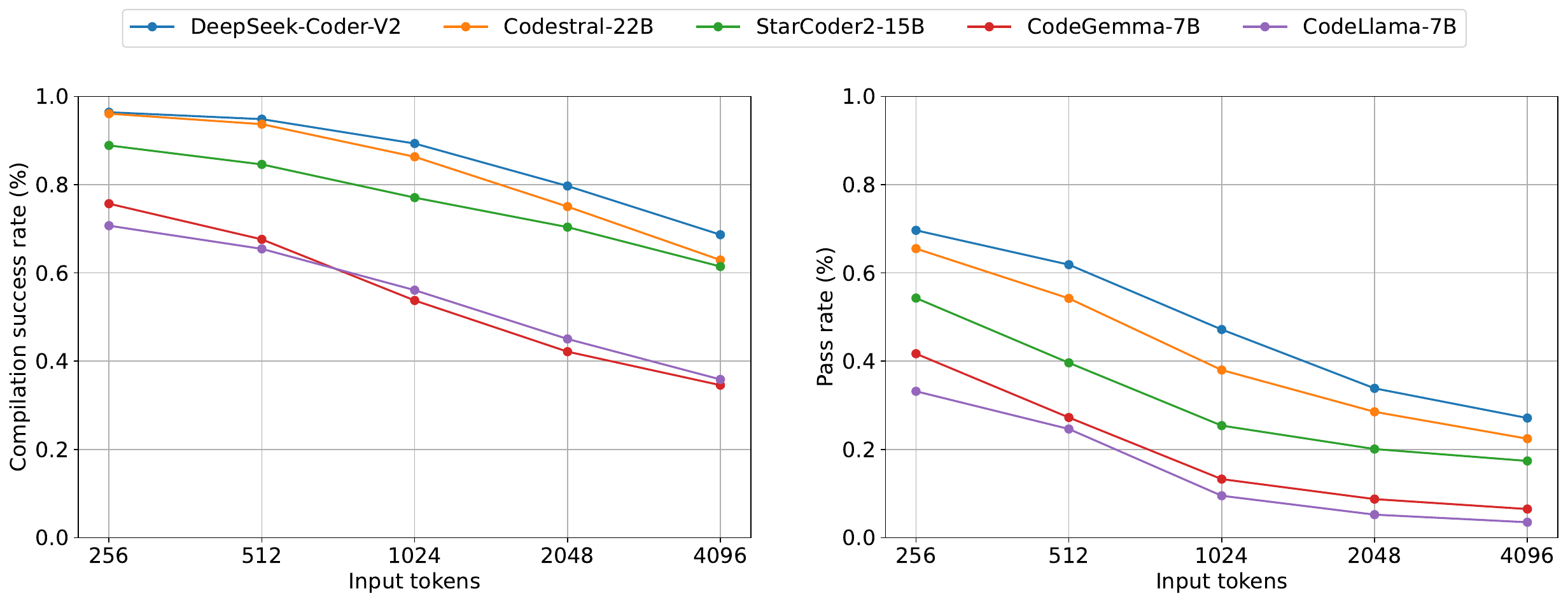}
    \caption{Compilation success rate and pass rate drop dramatically as input length increases.}
    \label{fig:pass_rate}
\end{figure}

As illustrated in \Cref{fig:pass_rate}, both compilation success rate and pass rate decline significantly as the number of input tokens increases. For smaller code LLMs, such as CodeLlama-7B~\cite{codellama}, the pass rate drops to nearly zero when the input code exceeds 1024 tokens. Even more capable models, despite being fine-tuned for longer sequences, experience a 50\% degradation in performance as the input code approaches 4096 tokens. This phenomenon suggests that, although code LLMs perform impressively on existing code translation benchmarks, they continue to face challenges in long code translation.
The challenge in long code translation stems from two key factors: code LLMs struggle to comprehend the full functionality of longer code inputs; second, they are more prone to mistakes when generating long code outputs. This issue persists across models of different sizes, indicating that long code translation is a common challenge for current code LLMs.

Take the most capable open-source code LLM (i.e., DeepSeek-Coder-V2~\cite{zhu2024deepseek}) as an example and delve into failed samples derived from DeepSeek-Coder-V2's translation results. Labeling failed samples with detailed causes is labor-intensive and impractical with large-scale analysis. Therefore, we leverage the extensive tests in our \textit{LongTrans} benchmark to estimate the quality and identify key characteristics of the failed samples. We execute all the translated programs on relevant tests and record their results. The failed samples are classified into five categories: compilation error, runtime error on all tests, runtime error on some tests, wrong answer on all tests, and wrong answer on some tests. This classification offers a novel perspective on evaluating failed examples, as the complexity increases progressively from passing compilation to executing without runtime errors, and finally, producing correct outputs.

\begin{figure}[ht]
    \centering
    \includegraphics[width=\textwidth]{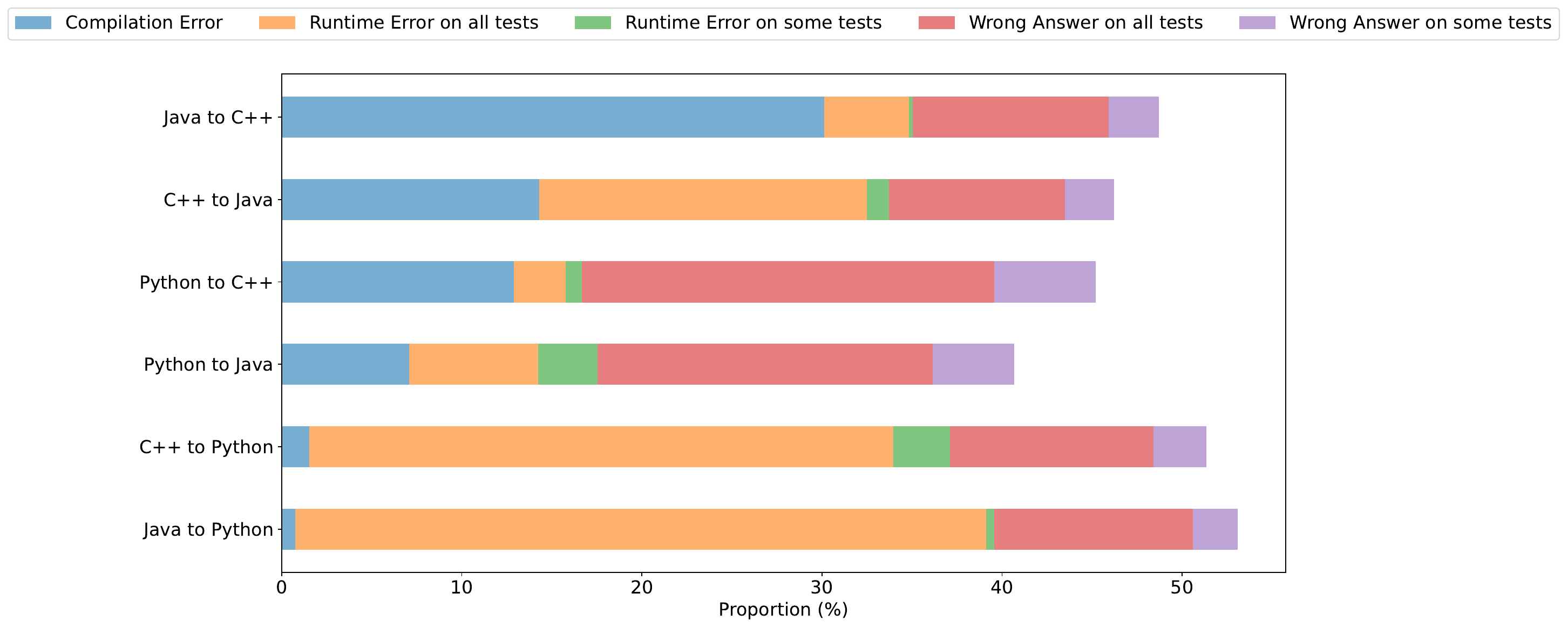}
    \caption{Distribution of failed samples across five categories in each translation dataset.}
    \label{fig:error_proportion}
\end{figure}

The distribution of failed samples across five categories in each translation dataset is illustrated in \Cref{fig:error_proportion}. As shown, the majority of failures vary across different translation datasets. For Python-to-Java and Python-to-C++ translations, wrong answers dominate the failure categories, with a notable proportion of wrong answers on all tests. 
In contrast, C++-to-Python and Java-to-Python translations exhibit a significantly higher occurrence of runtime errors. This observation aligns with intuition, as Python is a dynamically typed language with more lenient syntax checking, meaning many errors -- such as the absence of corresponding functions or type mismatches -- only surface during runtime. While examining C++-to-Java and Java-to-C++ translations, compilation errors emerge as the primary source of failure. 
In addition to the findings mentioned above, we have observed a significant discovery: whether it is a runtime error or a wrong answer, the translated program tends to fail on all tests in most cases. 
When a translated program fails on all tests, it suggests that the error is ``shallow'', meaning that the error is central to the program's logic or structure rather than specific to certain inputs or edges. Such errors often stem from incorrect syntax, missing functions, or major semantic misinterpretations that prevent any part of the translated program from working as intended. Because ``shallow'' errors prevent the program from passing any tests, they can be effectively exposed through executing the program on a set of tests.

\begin{figure}[htbp]
    \centering
    \includegraphics[width=\textwidth]{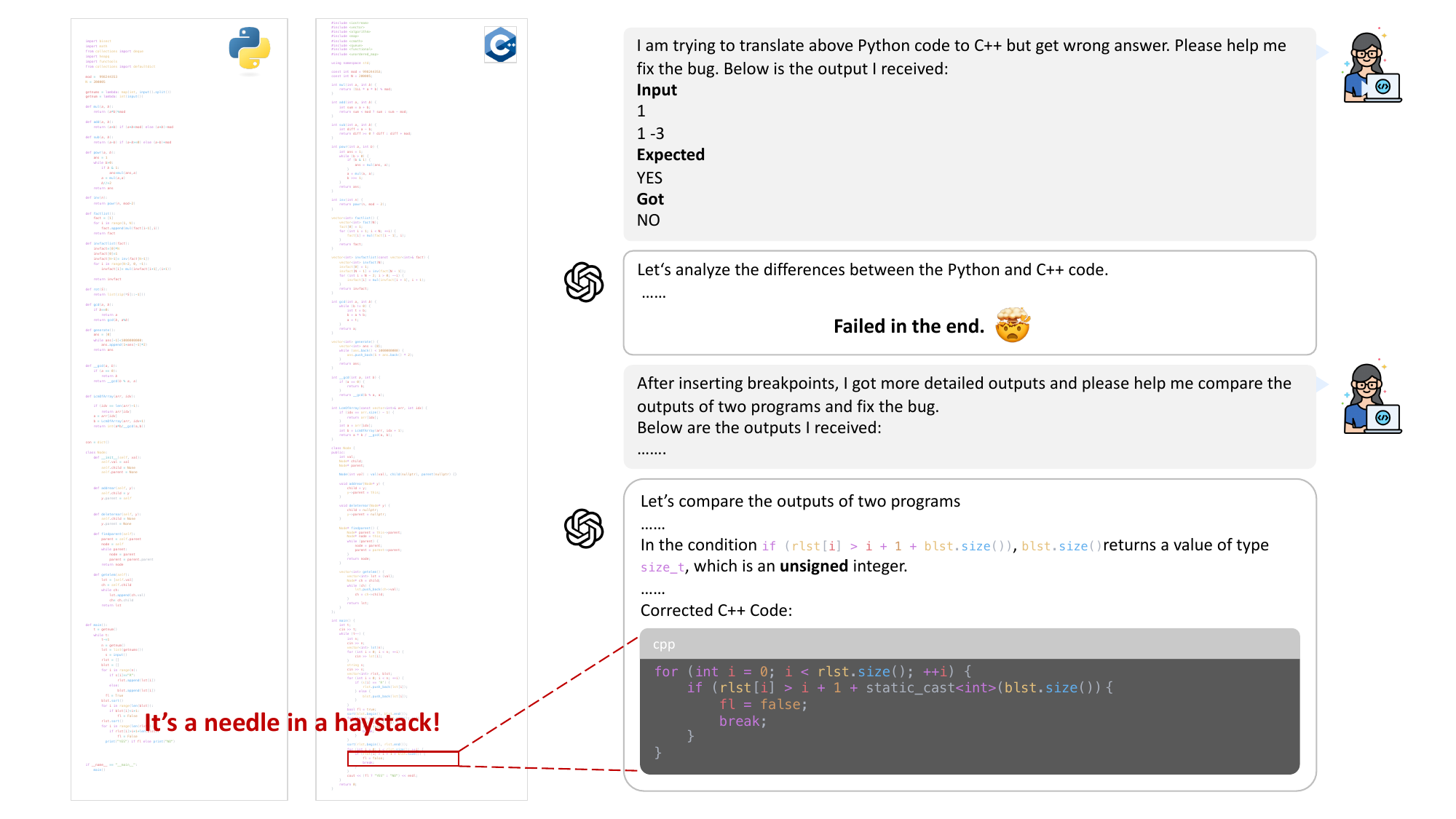}
    \caption{Code LLMs struggle to repair the translation errors hidden in hundreds of lines of code, but succeed when provided with detailed outputs of breakpoints.}
    \label{fig:motivation}
\end{figure}

However, identifying and repairing these errors remains a challenge, particularly in the context of translating long code. Recent work~\cite{Ni2024} suggests that LLMs lack a semantic understanding of how a program executes at run-time, leading to poor performance when instructed to repair the errors when provided with execution results from both the source and translated programs. As illustrated in \Cref{fig:motivation}, the Python program is translated to a C++ version but fails to pass the test. Unlike Python, where integers are unbounded, C++ has more strict type limitations. In the conditional statement \texttt{if (rlst[i] > i + 1 + blst.size())}, \texttt{blst.size()} returns a value of type \texttt{size\_t}, which is an unsigned integer. When comparing signed and unsigned integers, the C++ compiler converts the signed value to unsigned. This behavior leads to the semantic discrepancy between the Python and C++ programs when encountering negative values in the list.

This example shows that semantic differences between Python and C++ lead to incorrect results. If we could systematically identify these discrepancies and align the program states of the original and translated programs throughout execution, the final outputs would be guaranteed to match, thereby ensuring a correct translation. 
Inspired by this insight, we apply source code instrumentation -- a classical technique in dynamic program analysis -- to capture the program states and leverage code LLMs to align the program states from entry to exit. Further details will be provided in the next section.

\section{Method}

The proposed approach PAST consists of three stages: (1) the Instrumentation Stage, which leverages the code LLM to insert breakpoints into the source program in positions that will best reveal the execution flow and variable values, (2) the Translation Stage, which translates the source program into the target language while preserving the breakpoints. If the translated program fails on public tests, it proceeds to (3) the Program State Alignment Stage, where code LLMs are tasked with addressing inconsistencies between program states. In this stage, we employ two steps: direct repair and localize-re-translate. Initially, based on the error messages and the captured program states, we instruct the code LLM to repair the error directly. If this approach is unsuccessful, we localize the erroneous code snippet in the target program and its corresponding snippet in the source program, after that we re-translate the relevant source snippet.

\begin{algorithm}
\footnotesize
\caption{Translate with Instrumentation and Program State Alignment}
\label{algo:pipeline}
\KwIn{source program $x$, public\_tests $T$}
\KwOut{translated program}
$\hat{x} \gets \textnormal{instrumentation}(x)$ \Comment{{\color{blue}Stage 1: Instrumentation}} \\

$\hat{y} \gets \text{LLM\_translate}(\hat{x})$ \Comment{{\color{blue}Stage 2: Translation}} \\

\If{$\textnormal{check\_compilation\_error}(\hat{x}) \lor \textnormal{check\_compilation\_error}(\hat{y})$}{
    \Return $\textnormal{remove\_instrumentation}(\hat{y})$
}
$\textnormal{error}_y, \textnormal{output}_y \gets \textnormal{execute\_on\_tests}(\hat{y}, T)$ \\
$\textnormal{error}_x, \textnormal{output}_x \gets \textnormal{execute\_on\_tests}(\hat{x}, T)$ \\

\If{$\textnormal{output}_y = \textnormal{output}_x$}{
\Return $\textnormal{remove\_instrumentation}(\hat{y})$
}  
$\textnormal{diff}_{xy} \gets \textnormal{diff\_by\_line}(\textnormal{output}_y, \textnormal{output}_x)$\\\
$\hat{y}_{\textnormal{repaired}} \gets \textnormal{LLM\_direct\_repair}(\hat{y}, \textnormal{error}_y, \textnormal{diff}_{xy})$ \Comment{{\color{blue}Stage 3: Direct Repair}} \\
$\textnormal{error}_{\textnormal{repaired}}, \textnormal{output}_{\textnormal{repaired}} \gets \textnormal{execute\_on\_tests}(\hat{y}_{\textnormal{repaired}}, T)$ \\
\If{$\textnormal{output}_{\textnormal{repaired}} = \textnormal{output}_x$}{
    \Return $\textnormal{remove\_instrumentation}(\hat{y}_{\text{repaired}})$
}
$\hat{y}_{\textnormal{localize}} \gets \textnormal{LLM\_localize\_and\_retranslate}(\hat{y}, \textnormal{error}_y, \textnormal{diff}_{xy})$ \Comment{{\color{blue}Stage 3: Localize and Re-translate}} \\ 
$\textnormal{error}_{\textnormal{localize}}, \textnormal{output}_{\textnormal{localize}} \gets \textnormal{execute\_on\_tests}(\hat{y}_{\textnormal{localize}}, T)$ \\

\If{$\textnormal{output}_{\textnormal{localize}} = \textnormal{output}_x$}{
	\Return $\textnormal{remove\_instrumentation}(\hat{y}_{\textnormal{localize}})$
}
\Return $\textnormal{remove\_instrumentation}(\hat{y}_{\textnormal{localize}})$ \Comment{\color{blue} Failed  to Translate}
\end{algorithm}

As shown in \Cref{algo:pipeline}, our method aims to translate the source program $x$ into the target program $y$ while ensuring that both exhibit consistent behavior on public test cases. The process begins by instrumenting the source program $x$ to generate the instrumented program $\hat{x}$. A code LLM is then tasked with translating $\hat{x}$ into $\hat{y}$ in target language. If either $\hat{x}$ or $\hat{y}$ encounters a compilation error, the failed translation results are directly returned, although such cases are relatively rare. Next, $\hat{x}$ and $\hat{y}$ are executed on public test cases to obtain their error message (empty string if no error encountered) and outputs, which are then compared. If the outputs are identical, the instrumentation code in $\hat{y}$ is removed, and the result is returned. If the outputs differ, the line-by-line differences between the outputs are calculated. The LLM is then used to repair $\hat{x}$ based on these differences and the execution result type, producing the repaired program $\hat{y}_\text{repaired}$. The program is re-executed on public test cases, and the outputs are compared again. If the outputs are still different, the LLM is employed to localize and repair $\hat{y}$, producing $\hat{y}_\text{localize}$. If the outputs of $\hat{y}_\text{localize}$ and $\hat{x}$ remain inconsistent, the method returns the failed translation result.

\subsection{The Instrumentation Stage}
\label{subsection:breakpoint_insertion}
\begin{figure}[htbp]
    \centering
    \includegraphics[width=\textwidth]{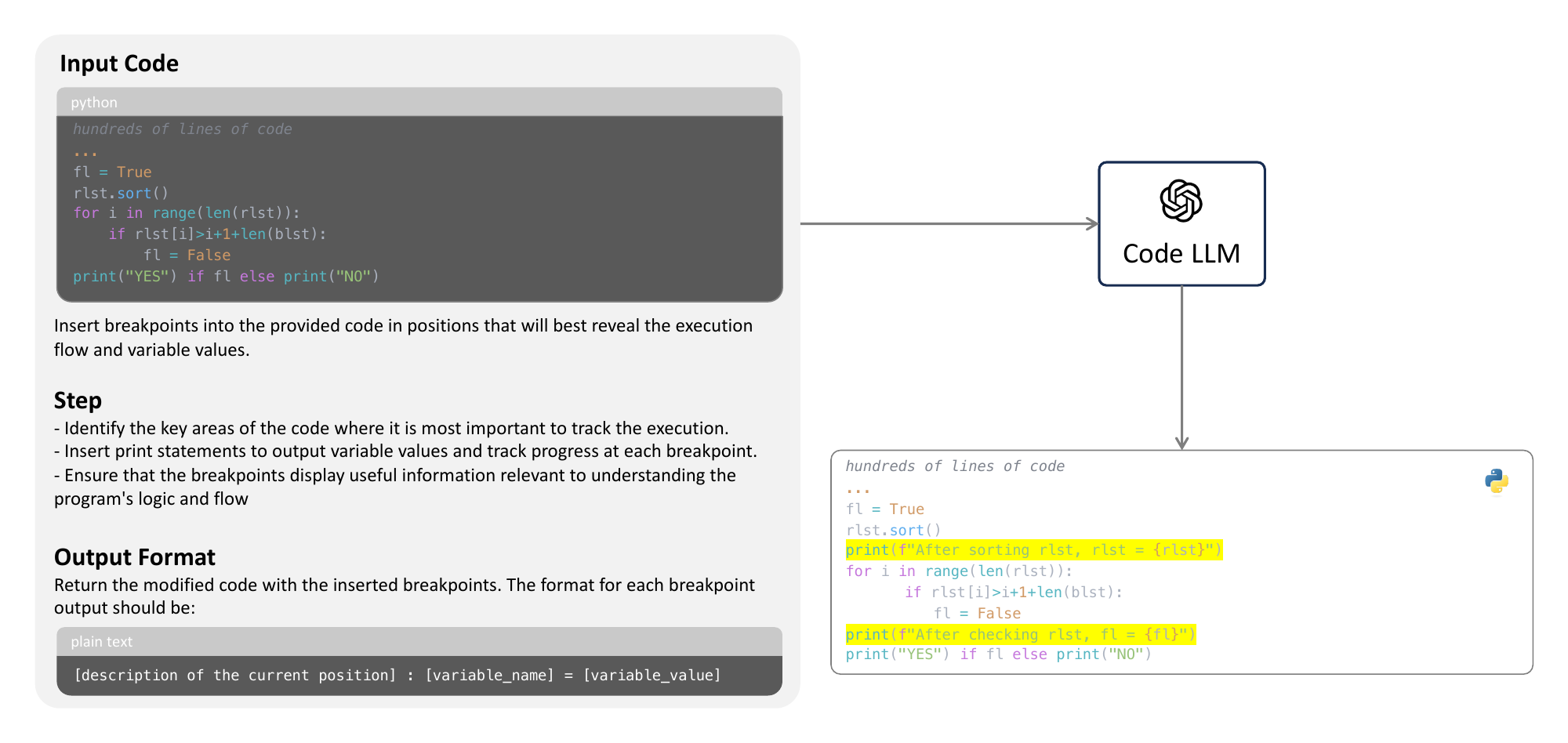}
    \caption{An Example of the Instrumentation Stage.}
    \label{fig:breakpoint}
\end{figure}

Instrumentation is widely employed to analyze program runtime states, and we leverage it here to capture and describe these states effectively. The process begins by inserting breakpoints at key points in the code to track the program's state. To accomplish this, we utilize the code LLM to identify the crucial positions and insert print statements to output variable values, and an example is illustrated in \Cref{fig:breakpoint}. To prevent an overload of overwhelming breakpoints and outputs, we emphasize that the information that breakpoints display should be relevant to understanding the program's logic and flow. Furthermore, to ensure that the breakpoint outputs can effectively help in identifying errors, we instruct the code LLM to format the breakpoint outputs as: "{[description of the current position]: [variable\_name] = [variable\_value]}". This format ensures the breakpoint outputs are clear and structured, facilitating more precise localization of program state inconsistencies in the Program State Alignment Stage.

\subsection{The Translation Stage}
\begin{figure}[htbp]
    \centering
    \includegraphics[width=\textwidth]{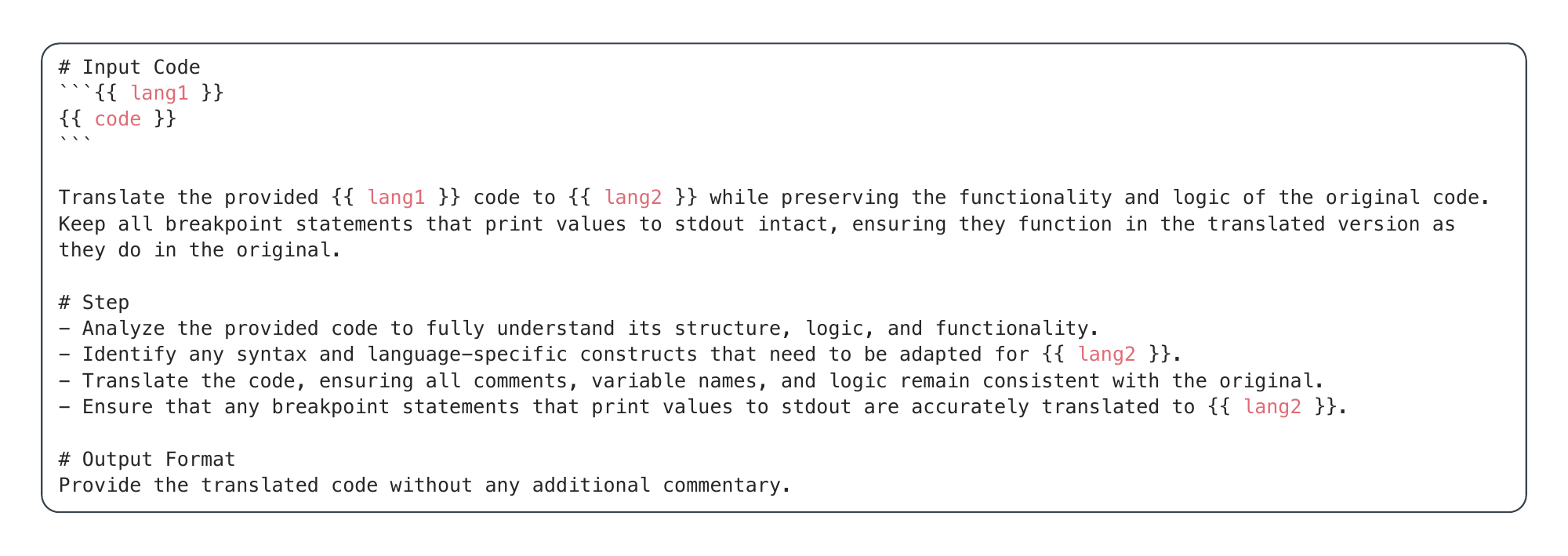}
    \caption{Prompt template for translating source program with breakpoints.}
    \label{fig:prompt_translate}
\end{figure}

After the Instrumentation Stage, the source program is now augmented with strategically placed breakpoints that facilitate  monitoring key variables and execution traces. In the Translation Stage, the primary task is to translate the source program into the target language while preserving the inserted breakpoints. The prompt can be formally defined using Jinja~\cite{JinjaDocs}, as shown in \Cref{fig:prompt_translate}, where \texttt{lang1} refers to the source language, \texttt{lang2} refers to the target language, and \texttt{code} represents the source program with breakpoints. With this prompt template, the code LLM can translate the source program while ensuring the semantics of the breakpoint statements remain unchanged.

\subsection{The Program State Alignment Stage}

\begin{figure}[htbp]
    \centering
    \includegraphics[width=\textwidth]{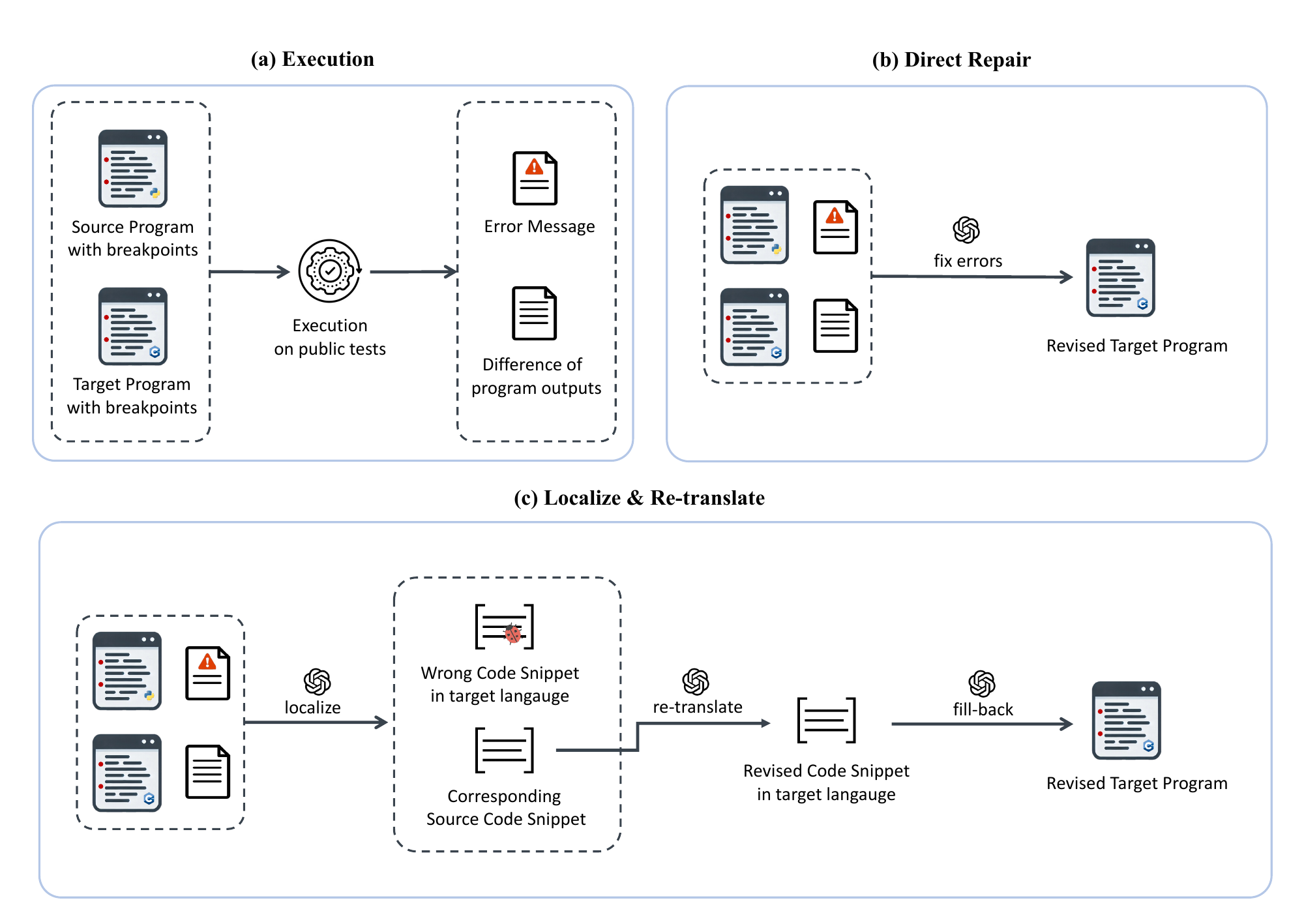}
    \caption{Illustration of the Program State Alignment Stage.}
    \label{fig:repair_stage}
\end{figure}

After the Translation Stage, we obtain the source and target programs both inserted with breakpoints. We then utilize execution to provide a comprehensive view of any semantic discrepancies between the source and target programs. This process involves executing both programs with identical test inputs and comparing their intermediate states and final outputs, as illustrated in the upper left corner of \Cref{fig:repair_stage}. The translation is considered successful if the target program produces intermediate states and outputs identical to those of the source program on public tests. If discrepancies are detected, we proceed to the next step, where two repair steps are employed to address the inconsistencies between the original and translated programs.

The first repair step is direct repair. During execution, the program generates an error message alongside its output, which may indicate a compilation error, runtime error, or wrong answer. The program output includes both breakpoint outputs, formatted as described in \Cref{subsection:breakpoint_insertion}, and the final output. The breakpoint outputs include descriptions of the positions and variable values, allowing us to identify inconsistencies in the program states as soon as the breakpoint outputs of both programs diverge. We employ text diff tools to compare the program outputs, highlighting the difference and minimizing redundant, irrelevant content to better focus the code LLM's attention. The source program, target program, error message, and output differences are provided to the code LLM, which is tasked with aligning program states based on this information. While direct repair appears intuitive, it has proven to be highly effective when supplemented with breakpoint outputs. We will demonstrate the effectiveness of direct repair in our experiments.

The second repair step is localize-and-re-translate. This step is motivated by the observation that code LLMs tend to struggle with translating entirely programs accurately but excel when focused on shorter code snippets. Therefore, we first direct the code LLM to localize the translation error and extract the erroneous code snippet in the target language, alongside its corresponding snippet in the source language -- referred to as the target code snippet and source code snippet, respectively. We then re-translate the source code snippet and reintegrate the new target snippet back into the target program. By eliminating the need to distribute attention across lengthy input, the code LLM can better address subtle semantic inconsistencies. Consequently, localize-and-re-translate can further enhance translation accuracy beyond what is achieved through direct repair. 

Finally, the translated program will executed on private tests to verify that it produces outputs identical to those of the original program. The private tests included in the benchmark are designed to be as exhaustive as possible, ensuring that translated programs cannot be incorrectly marked as correct due to insufficient test coverage. This rigorous testing approach helps maintain the integrity of the evaluation process.

\section{Benchmark}

In this section, we provide detailed descriptions of our proposed \textit{LongTrans} benchmark, how it was collected and processed, and its comparison with other code translation benchmarks. The \textit{LongTrans} benchmark consists of programs from three popular programming languages: C++, Java, and Python, and utilizes sufficient tests to evaluate translation results based on execution results. The detailed statistics about the \textit{LongTrans} dataset are presented in \Cref{table:statistics}. 

\begin{table}[htbp]
\centering
\footnotesize
\caption{Statistics for the \textit{LongTrans} benchmark. The number of tokens is based on the RoBERTa~\cite{Liu2019} tokenizer. Length is the number of characters.}

\begin{tabular}{cccccc}
\toprule
\textbf{Split}    & \textbf{Language} & \textbf{Samples} & \textbf{Avg. Tokens per Sample} & \textbf{Avg. Length} & \textbf{Avg. Tests per Sample} \\ \midrule
\multirow{3}{*}{Train} & C++      & 96791   &      775    &    1465    & 
99\\
                       & Java     & 62994   &      1919   &    3578    & 112                           \\
                       & Python   & 58517   &      429    &    785     & 116                       \\ \midrule
\multirow{3}{*}{Valid} & C++      & 12734   &      793    &    1519    & 202                       \\
                       & Java     & 8251    &      2155   &    3919    & 198                      \\
                       & Python   & 8396    &      711    &    1294    & 199                       \\ \midrule
\multirow{3}{*}{Test}  & C++      & 14677   &      825    &    1585    & 201                      \\
                       & Java     & 8894    &      2379   &    4342    & 197                      \\
                       & Python   & 8248    &      619    &    1147    & 197                      \\ \bottomrule
\end{tabular}

\label{table:statistics}
\end{table}

\subsection{Data Collection and Processing}

Our dataset was collected from the CodeContests dataset~\cite{Li2022}, which consists of programming problems collected from Codeforces, LeetCode, and other online judge platforms. In addition to the problems themselves, the dataset includes corresponding solutions and high-quality test cases. This enables us to evaluate translated programs through execution instead of relying on lexical similarity, which has been proven to be a poor proxy for functional equivalence~\cite{Chen2021}.

By collecting submitted solutions from the CodeContests dataset, we obtained adequate samples written in the most popular submission languages: C++, Python, and Java. However, we identified a significant number of false positive samples, where incorrect submissions are erroneously marked as correct due to insufficient test coverage. To mitigate this issue, we exclusively select samples collected from Codeforces, where the average number of tests per problem is notably higher than other sources. Moreover, we observed that some programming problems had many identical or nearly identical submissions, leading to an abundance of near-duplicate samples. To address this, we employed the MinHash~\cite{Broder1997} algorithm for deduplication, effectively reducing redundancy. Furthermore, we performed clustering within each problem's submissions to identify and select representative samples. This approach minimized the impact of near-duplicate submissions on the overall quality and diversity of the benchmark.

\begin{table}[htbp]
\centering
\caption{Comparsion of \textit{LongTrans} with existing code translation datasets. We report Samples and Avg. Tokens per Sample for Train/Valid/Test sets of each dataset. For Samples, we report \textbf{program-level} counts for AVATAR, CoST, and XLCoST. }
\resizebox{\textwidth}{!}{
\begin{tabular}{cccccc}
\toprule
\textbf{Dataset} & \textbf{Unit Tests} & \textbf{Languages} & \textbf{Samples} & \textbf{Avg. Tokens per Sample} & \textbf{Avg.Length} \\ \midrule
CodeXGLUE~\cite{Lu2021}   &  \ding{55}   & Java, C\#          & 10300 / 500 / 1000              & 59 / 63 / 58                  & 205 / 218 / 202     \\
AVATAR~\cite{Ahmad2023}    & \Checkmark         & Java, Python       & 7113 / 476 / 1906               & 239 / 235 / 234               & 691 / 687 / 688     \\
TransCoder~\cite{transcoder}    &   \Checkmark   & C++, Java, Python  & -- / 470 / 948                & -- / 119 / 120                 & - / 313 / 311       \\
CoST~\cite{must}        &   \ding{55}     & 7 languages        & 14260 / 1029 / 1449             & 272 / 180 / 199               & 770 / 458 / 511     \\
XLCoST~\cite{Zhu2022}    &   \ding{55}       & 7 languages        & 102559 / 5892 /10632            & 234 / 232 / 222               & 644 / 634 / 606     \\
MultilingualTrans~\cite{Yan2023}  & \ding{55}   & 9 languages        & 19115 / 3759 / 7545              & 398 / 421 / 491               & 1099 / 1135 / 1358  \\
NicheTrans~\cite{Yan2023}     & \ding{55}   & 37 languages       & 165457 / 23509 / 47502             & 292 / 375 / 505               & 785 / 995 / 1372    \\
LLMTrans~\cite{Yan2023}     &   \Checkmark    & 8 languages        & -- / -- / 350                & -- / -- / 270                   & - / - / 745         \\
CoTran~\cite{Jana2023}     &  \Checkmark       & Java, Python       & 55178 / 442 / 1745              & 288 / 242 / 249               & 619 / 639 / 654     \\ \midrule
\textit{LongTrans}      &  \Checkmark    & C++, Java, Python  & 218302 / 29381 / 31819            & 1012 / 1152 / 1206            & 1893 / 2129 / 2243   \\ \bottomrule
\end{tabular}
}

\label{table:comparsion}
\end{table}

\subsection{Benchmark Comparisons}
As indicated by its name, \textit{LongTrans} is distinguished by the average length of its samples. As shown in table \ref{table:comparsion}, the average number of tokens per sample in \textit{LongTrans} is approximately 2-3 times longer than in previous benchmarks, presenting a significant challenge for code LLMs. The unit tests mean whether or not to run the generated code when testing, for example, computing the BLEU metric does not require execution, but computing the accuracy does.

\section{Experiment Setting}
In this section, we will describe our experimental setup, which consists of the studied large language models, evaluation metrics, implementation of the experiment, and research questions.

\subsection{Studied Models}
\label{models}
We introduce different families of recent LLMs of diverse sizes for empirical study. Their detailed information is listed below.
\begin{itemize}
 \item \textbf{CodeGeeX4~\cite{zheng2023codegeex}}: is the open-source version of the latest CodeGeeX4 model series. It is a multilingual code generation model continually trained on the GLM-4~\cite{GLM2024}, significantly enhancing its code generation capabilities. It is developed upon the ChatGLM2~\cite{DuQLDQY022} architecture and is enhanced with an extensive dataset of coding examples.
 \item \textbf{CodeGemma~\cite{team2024codegemma}}: is a collection of specialized open code models built on top of Gemma, capable of a variety of code and natural language generation tasks. CodeGemma 7B pretrained (PT) and instruction-tuned (IT) variants have remarkably resilient natural language understanding, excel in mathematical reasoning, and match the code capabilities of other open-source models. 

 \item \textbf{CodeQwen1.5~\footnote{https://huggingface.co/Qwen/CodeQwen1.5-7B}}: is the Code-Specific version of Qwen1.5~\cite{Bai2023}. It is a transformer-based decoder-only language model pre-trained on a large amount of data of codes. CodeQwen1.5 is trained on 3 trillion tokens of data of codes, and it includes group query attention~\cite{Ainslie2023} to reduce KV cache and acclerate inference.
 \item \textbf{Qwen2.5~\cite{qwen2}}: is the latest series of Qwen large language models. Qwen2.5 is pre-trained on their large-scale dataset, encompassing up to 18 trillion tokens. It has acquired significantly more knowledge and has greatly improved capabilities in coding and mathematics compared to its preceding series.
 \item \textbf{StarCoder2~\cite{Lozhkov2024}}: is a publicly accessible model with a substantial parameter count of 15 billion. It is pre-trained on over 600 programming languages from The Stack v2, as well as natural language text from sources like Wikipedia, Arxiv, and GitHub issues, ensuring proficiency across a wide range of coding tasks.
 \item \textbf{Codestral\footnote{https://mistral.ai/news/codestral/}}: is trained on a diverse dataset of 80+ programming languages, including the most popular ones, such as Python, Java, C, C++, JavaScript, and Bash. It also performs well on more specific ones like Swift and Fortran. This broad language base ensures that Codestral can assist developers in various coding environments and projects.

 \item \textbf{Mistral-Large-2~\footnote{https://mistral.ai/news/mistral-large-2407/}}: is the new generation of flagship model developed by Mistral, boasting 123 billion parameters. Compared to its predecessor, Mistral Large 2 is significantly more capable in code generation, mathematics, and reasoning.
 \item \textbf{DeepSeek-Coder~\cite{guo2024deepseek}}: is a range of open-source code models with sizes from 1.3B to 33B. It is pre-trained from scratch on 2 trillion tokens, with a composition of 87\% code and 13\% natural language in both English and Chinese. For coding capabilities, DeepSeek-Coder achieves state-of-the-art performance among open-source code models on multiple programming languages and various benchmarks.
 
 \item \textbf{DeepSeek-Coder-V2~\cite{zhu2024deepseek}}: is an open-source Mixture-of-Experts (MoE) code language model that achieves performance comparable to GPT-4 in code-specific tasks. Built on the DeepSeekMoE~\cite{Dai2024} framework, DeepSeek-Coder-V2 comprises 236 billion parameters, with 21 billion activated during inference.
 \item \textbf{LlaMA-3.1~\cite{touvron2023llama}}: A family of multilingual large language models ranging from 8B to 405B. We include two versions of 8B/70B of LlaMA-3.1 for experiments. The Llama 3.1 instruction-tuned, text-only models are optimized for multilingual dialogue and excel on common industry benchmarks, outperforming many open-source and closed chat models.
 \item \textbf{CodeLlaMA~\cite{codellama}}:  encompasses a series of code-centric Large Language Models (LLMs) that are derivatives of LlaMA2~\cite{Hugo:2023llama2}. Available in three sizes — 7B, 13B, and 34B — these models undergo continued training on a vast 500 billion token code corpus, building upon the foundational LlaMA2 architecture.
\item  \textbf{GPT-4o and GPT-4o-mini~\cite{gpt4}} are advanced generative AI models developed by OpenAI. While they are not explicitly trained for code generation, they also demonstrate notable performance in this domain. Their effectiveness in handling code generation tasks is largely attributed to their massive scale in terms of parameter count. 
\end{itemize}

\subsection{Evaluation Metrics}
\label{subsection:evaluation_metrics}
Following the previous studies in the code translation field~\cite{transcoder,transcoder2,transcoder3,yang2024exploring}, we adopt Computational Accuracy (CA) to assess the performance of each model.

The computational accuracy computes the ratio that the translated codes can produce the same execution results as the ground truth source code while giving the same inputs~\cite{yang2024exploring}. This metric measures the semantic consistency of two programs, assuming that the given example covers all boundary conditions, which can be formalized as:

\begin{equation}
    CA = \frac{\sum_{k=1}^N ca(y_k, \hat{y_k})}{N}~\text{, where }
\end{equation}
\begin{equation}
ca(y_k, \hat{y_k}) =\left\{
\begin{aligned}
1 & & Exec_k(y_k) = Exec_k(\hat{y_k}) \\
0 & & Exec_k(y_k) \neq Exec_k(\hat{y_k})~\text{.}
\end{aligned}
\right.
\end{equation}

$N$ denotes the total number of translation samples, $y_k$ denotes the ground truth of the $k$-th sample, and $\hat{y_k}$ denotes the translated program via a certain transpiler for the $k$-th sample. $Exec_k(\cdot)$ denotes the execution
result of a program with the test suite of the $k$-th sample. For example, even if $y_k$ and $\hat{y_k}$ are not
identical literally, they are considered a correct translation in CA, as long as $Exec_k(y_k)$ = $Exec_k(\hat{y_k})$.

\subsection{Implementation}
We access GPT-4o and GPT-4o-mini via OpenAI's API\footnote{https://platform.openai.com/}, while for open-source code LLMs, we load their weights from Huggingface\footnote{https://huggingface.co/} and perform inference using vllm~\cite{Kwon2023}, which effectively accelerates the inference with paged attention. To ensure fairness, we maintain consistent default settings for all code LLMs in the empirical study. We use nucleus sampling~\cite{Holtzman2019} with parameters set to $top\_p=0.95, temperature=0.2$. We adopt a zero-shot learning setting, as the code LLMs we investigated are instruction fine-tuned~\cite{Ouyang2022}, and this setting closely aligns with real-world scenarios. For our approach, the sampling settings of backbone LLMs in three stages are kept as default.

\subsection{Research Questions and Evaluation Methodology}

In the next section, we will introduce five research questions of this work and show the evaluation results.

\textbf{RQ1: What is the performance of recent LLMs at different scales in long code translation in terms of computation accuracy?} 
This RQ extensively assesses the code translation performance of recent LLMs. In this empirical study, we evaluate the models mentioned in Section~\ref{models} on six translation datasets of \textit{LongTrans}, i.e., C++ to Python, Python to C++, Java to C++, C++ to Java, Java to Python, and Python to Java. In those datasets, we quantitatively explore the practicality of LLMs in the long code translation task.

\textbf{RQ2: How does our approach perform with different LLMs?}
This RQ aims to examine the
effectiveness of our approach and its generalization capability on various LLMs. In experiments, the DeepSeekCoder-V2, Qwen2.5, Mistral-Large-2, GPT-4o-mini, and the state-of-the-art LLM, GPT-4o are selected for investigation.

\textbf{RQ3: How effective are the different components of the method?}
Beyond translation, our approach consists of three components to boost the code translation, including instrumentation stage, direct-repair step and localize \& re-translation step. The ablation study of the different components is designed to investigate their contributions in terms of computation accuracy on five LLMs including DeepSeekCoder-V2, Qwen2.5,  Mistral-Large-2, GPT-4o-mini, and GPT-4o.

\textbf{RQ4: How about LLM translation performance as code length increases?}

In this RQ, we investigate the impact of code length on the translation performance of code LLMs, specifically analyzing how variations in the length of input code affect the computational accuracy of the generated outputs. We conduct experiments on six translation datasets from the \textit{LongTrans} benchmark, comparing the performance of code LLMs and their enhanced version with our approach, and present the average results. These datasets are categorized by source code length into five distinct ranges for evaluation: less than 256 tokens, 256-512 tokens, 512-1024 tokens, 1024-2048 tokens, and more than 2048 tokens.

\textbf{RQ5: How about the sensitivity of the approach to the quality of the public tests?}
The quality of public tests impacts the efficiency of code LLMs, as tests with low coverage may not reveal potential issues. To further demonstrate the generality of our approach, we examine its sensitivity to the quality of the public tests. By varying the quality of public tests to adjust the executed feedback quality, we aim to explore whether our approach remains effective even when the executed feedback quality is of lower quality.

\begin{table}[ht]
\centering
\small
\caption{The empirical results of LLMs across different programming languages.}
\resizebox{\textwidth}{!}{
\begin{tabular}{lccccccc}
\toprule
\multirow{2}{*}{\textbf{Model}} & \multicolumn{6}{c}{\textbf{Language}} & \multirow{2}{*}{\textbf{Average}} \\ \cmidrule{2-7}
  & C++-Java & C++-Python & Java-C++ & Java-Python & Python-C++ & Python-Java & \\
  \midrule
CodeGeeX4-9B          & 35.77\% & 28.09\% & 35.37\% & 40.09\% & 37.52\% & 38.78\% & 35.94\% \\ 
CodeGemma-7B          & 14.90\% & 19.46\% & 31.70\% & 20.67\% & 26.58\% & 27.00\% & 23.39\% \\ 
CodeQwen1.5-7B    & 24.36\% & 21.13\% & 31.25\% & 23.71\% & 31.64\% & 32.45\% & 27.42\% \\ 
StarCoder2-15B & 30.39\% & 26.67\% & 46.34\% & 36.72\% & 37.63\% & 34.41\% & 35.36\% \\ 
Codestral-22B          & 43.59\% & 38.16\% & 47.70\% & 41.80\% & 51.21\% & 53.07\% & 45.92\% \\ 
Mistral-Large-2-123B & 48.31\% & 46.18\% & 54.47\% & 58.11\% & 50.73\% & 46.25\% & 50.68\% \\
\midrule
CodeLlama-7B & 18.34\% & 12.46\% & 24.25\% & 15.95\% & 13.65\% & 13.40\% & 16.34\% \\ 
CodeLlama-13B & 17.51\% & 18.21\% & 26.50\% & 19.42\% & 15.65\% & 15.17\% & 18.74\% \\ 
CodeLlama-34B & 21.19\% & 11.79\% & 21.87\% & 6.91\% & 17.70\% & 15.48\% & 15.82\% \\ 
\midrule
DeepSeekCoder-1.3B  & 21.47\% & 15.45\% & 30.02\% & 18.19\% & 24.19\% & 15.04\% & 20.73\% \\ 
DeepSeekCoder-6.7B  & 40.03\% & 27.47\% & 42.91\% & 34.13\% & 46.19\% & 45.26\% & 39.33\% \\ 
DeepSeekCoder-33B  & 49.61\% & 40.10\% & \textbf{63.90\%} & 52.77\% & 54.46\% & 52.71\% & 52.26\% \\ 
DeepSeekCoder-V2  & 53.78\% & 48.64\% & 51.28\% & 46.89\% & 54.81\% & \textbf{59.31\%} & 52.45\% \\ 
\midrule
Llama3.1-8B            & 32.62\% & 27.36\% & 29.08\% & 30.33\% & 35.26\% & 37.21\% & 31.98\% \\ 
Llama3.1-70B           & 50.18\% & 43.70\% & 49.71\% & 47.72\% & 53.30\% & 55.42\% & 50.01\% \\ 
\midrule
GPT-4o-mini & 45.66\% & 45.11\% & 62.29\% & \textbf{61.70\%} & 52.06\% & 46.71\% & 52.26\% \\
GPT-4o & \textbf{59.18\%} & \textbf{50.10\%} & 62.17\% & 60.99\% & \textbf{56.93\%} & 55.71\% & \textbf{57.51\%} \\
\midrule
Average & 35.70\% & 30.59\% &	41.81\% &	36.24\% &	38.79\% &	37.85\% &	36.83\% \\
\bottomrule
\end{tabular}
}
\label{sota}
\begin{tablenotes}
\item[$\Phi$] \scriptsize For clarification, we adopt ``-'' to concatenate LLM and each of its corresponding parameter sizes for discrimination. For example, Llama3.1 with 70B parameters is dubbed Llama3.1-70B. An exception is that DeepSeek-Coder-V2 has 236B parameters in total, with 21B activated during inference. Bold indicates the optimal value on the current dataset. 
\end{tablenotes}
\end{table}

\section{Experiment Results}
\subsection{RQ1: What is the performance of recent LLMs at different scales in long code translation in terms of computation accuracy?}
\Cref{sota} presents the empirical results of each model on the \textit{LongTrans} benchmark, with the performance measured in terms of Correctness Accuracy. (1) As the results clearly indicate, GPT-4o consistently performs the best on most translation datasets, achieving accuracy rates of 59.18\%, 50.10\%, 61.70\%, and 56.93\% in the C++ to Java, C++ to Python, Java to Python, and Python to C++ datasets. Additionally, it is surprising that DeepSeek-Coder outperforms GPT-4o in the Java to C++ and Python to Java datasets. (2) Overall, as the model parameter size increases, the translation performance of large language models gradually improves. (3) In long code translation scenarios, large language models exhibit better average performance when translating from Java to C++. This may be due to the structural similarities between Java and C++, making the mappings more explicit in long code translations. In contrast, the average performance for translations from C++ to Python is relatively lower, likely because of the significant syntactic and semantic differences between C++ and Python. These differences lead to unclear structural correspondences, particularly in long code translations, which can result in mistranslations and omissions.

\begin{boxK}

\small \faIcon{lightbulb} \textbf{Answer to RQ1:} 
The experimental results demonstrate that larger model sizes correlate with improved translation accuracy, especially in long code translation. Models also translate Java to C++ more effectively, likely due to structural similarities, whereas C++ to Python translations suffer from syntactic and semantic differences, leading to more frequent errors and omissions.
\end{boxK}

\vspace{-0.3cm}
\begin{table}[!ht]
\centering
\footnotesize
\caption{Comparison of the performance of LLMs of code with and without PAST.}
\begin{tabular}{cccccccc}
\toprule
\multirow{2}{*}{Model} & \multicolumn{2}{c}{C++} & \multicolumn{2}{c}{Java} & \multicolumn{2}{c}{Python} & \multirow{2}{*}{Average} \\
\cline{2-3} \cline{4-5} \cline{6-7}
& Java & Python & C++ & Python & C++ & Java & \\
\midrule
DeepSeekCoder-V2 & 53.78\% & 48.64\% & 51.28\% & 46.89\% & 54.81\% & 59.31\% & 52.45\% \\
+ PAST & 78.10\% & 74.27\% & 74.12\% & 80.92\% & 71.83\% & 79.33\% & 76.43\% \\ 
\rowcolor{gray!20}
\textit{Improvement} & +45.22\%	& +52.69\%	& +44.54\%	& +72.57\% &	+31.05\% & 	+33.75\%	& +46.64\%
 \\
\midrule
Qwen2.5-72B & 38.52\% & 39.25\% & 42.33\% & 51.40\% & 50.37\% & 53.69\% & 45.93\% \\
+ PAST & 66.34\% & 63.52\% & 55.17\% & 72.83\% & 69.26\% & 82.84\% & 68.49\% \\
\rowcolor{gray!20}
\textit{Improvement} & +72.22\%	& +61.83\% & +30.33\% & +41.69\% & +37.50\% & +54.29\% & +49.65\%
 \\
\midrule
Mistral-Large-2-123B & 48.31\% & 46.18\% & 54.47\% & 58.11\% & 50.73\% & 46.25\% & 50.68\% \\
+ PAST & 72.13\% & 65.80\% & 65.05\% & 68.42\% & 70.78\% & 76.00\% & 69.67\% \\
\rowcolor{gray!20}
\textit{Improvement} & +49.31\% & +42.49\% & +19.42\% & +17.74\% & +39.52\% & +64.32\% & +38.80\% \\
\midrule
GPT-4o-mini & 45.66\% & 45.11\% & 62.29\% & 61.70\% & 52.06\% & 46.71\% & 52.26\% \\
+ PAST & 70.02\% & 59.26\% & 78.76\% & 72.69\% & 82.78\% & 85.93\% & 74.91\% \\
\rowcolor{gray!20}
\textit{Improvement} & +53.35\% & +31.37\% & +26.44\% & +17.81\% & +59.01\% & +83.96\% & +45.32\% \\
\midrule
GPT-4o & 59.18\% & 50.10\% & 62.17\% & 60.99\% & 56.93\% & 55.71\% & 57.51\% \\
+ PAST & 86.92\% & 78.13\% & 81.79\% & 82.92\% & 86.42\% & 92.24\% & 84.70\% \\
\rowcolor{gray!20}
\textit{Improvement} & +46.87\% & +55.95\% & +31.56\% & +35.96\% & +51.80\% & +65.57\% & +47.95\% \\

\bottomrule
\end{tabular}
\label{exp2}
\end{table}
\vspace{-0.3cm}

\subsection{RQ2: How does our approach perform with different LLMs?}
In order to verify the general applicability of our method, we conducted validation on multiple LLMs with different architectures and different parameters, and the results are shown in \Cref{exp2}. As can be seen, our approach consistently boosts the performance of LLMs, including DeepSeekCoder-V2, Qwen2.5, Mistral-Large-2, GPT-4o-mini, and GPT-4o, of 46.64\%, 49.65\%, 38.80\%, 45.32\%, and 47.95\% improvement in terms of computation accuracy on average. This demonstrates the effectiveness of our approach for improving long code translation of LLMs with different architectures and different parameters. 
Particularly, \Cref{exp2} demonstrates the translation on GPT-4o achieves the computation accuracy of 84.70\% on average, which is the highest among all experimented LLMs. 


\begin{boxK}
\small \faIcon{lightbulb} \textbf{Answer to RQ2:} Our method consistently improves the long code translation performance across various large language models with different architectures and parameters. This confirms the effectiveness of our approach for enhancing long code translation in diverse LLMs.
\end{boxK}

\subsection{RQ3: How effective are the different components of the method?}

As shown in \Cref{tableablation}, the ablation study is conducted on three important components, instrumentation stage, direct-repair step, and localize \& re-translation step of the PAST.

\begin{table}[h!]
\centering
\footnotesize
\caption{Model performance across different programming languages and improvements with techniques such as direct repair (\textit{R}), instrumentation (\textit{I}), and localization \& re-translation (\textit{L\&R}).}
\begin{tabular}{lccccccc}
\toprule
\multirow{2}{*}{Model} & \multicolumn{2}{c}{C++} & \multicolumn{2}{c}{Java} &\multicolumn{2}{c}{Python} & \multirow{2}{*}{Average} \\ \cline{2-3} \cline{4-5} \cline{6-7} 
                       & Java   & Python & C++ & Python   & C++ & Java \\   \midrule
DeepSeekCoder-V2       & 53.78\% & 48.64\% & 51.28\% & 46.89\% & 54.81\% & 59.31\%  & 52.45\%\\  
+ \textit{R}               & 59.57\% & 54.16\% & 53.64\% & 49.65\% & 58.06\% & 60.43\%  & 56.48\% \\  
+ \textit{R} + \textit{I}          & 75.62\% & 72.09\% & 73.02\% & 77.56\% & 70.40\% & 77.49\%  & 74.36\% \\  
+ \textit{R} + \textit{I} + \textit{L\&R} & 78.10\% & 74.27\% & 74.12\% & 80.92\% & 71.83\% & 79.33\%  & 76.43\% \\  \midrule
Qwen2.5-72B            & 38.52\% & 39.25\% & 42.33\% & 51.40\% & 50.37\% & 53.69\%  & 45.93\% \\ 
+ \textit{R} & 47.80\% & 45.50\% & 43.14\% & 56.67\% & 54.03\% & 57.84\% & 50.83\% \\
+ \textit{R} + \textit{I}                & 63.33\% & 60.80\% & 51.05\% & 68.55\% & 66.84\% & 80.08\%  & 65.11\% \\  
+ \textit{R} + \textit{I} + \textit{L\&R}         & 66.34\% & 63.52\% & 55.17\% & 72.83\% & 69.26\% & 82.84\%  & 68.33\% \\  \midrule
Mistral-Large-2-123B        & 48.31\% & 46.18\% & 54.47\% & 58.11\% & 50.73\% & 46.25\%  &50.68\% \\  
+ \textit{R} & 57.35\% & 50.79\% & 55.92\% & 59.57\% & 56.30\% & 48.25\% & 54.70\%  \\
+ \textit{R} + \textit{I}                 & 70.06\% & 63.84\% & 61.02\% & 66.29\% & 68.85\% & 73.57\%  & 67.27\% \\  
+ \textit{R} + \textit{I} + \textit{L\&R}         & 72.13\% & 65.80\% & 65.05\% & 68.42\% & 70.78\% & 76.00\%  & 69.70\%\\  \midrule 
GPT-4o-mini            & 45.66\% & 45.11\% & 62.29\% & 61.70\% & 52.06\% & 46.71\% & 52.26\%\\  
+ \textit{R} & 53.94\% & 46.84\% & 66.09\% & 63.06\% & 60.89\% & 49.79\% &56.77\% \\
+ \textit{R} + \textit{I}                & 67.30\% & 56.88\% & 77.64\% & 70.56\% & 78.90\% & 84.24\% & 72.59\% \\  
+ \textit{R} + \textit{I} + \textit{L\&R}          & 70.02\% & 59.26\% & 78.76\% & 72.69\% & 82.78\% & 85.93\%  & 74.91\% \\  \midrule
GPT-4o                 & 59.18\% & 50.10\% & 62.17\% & 60.99\% & 56.93\% & 55.71\%  & 57.51\% \\ 
+ \textit{R} & 66.16\% & 58.09\% & 65.52\% & 62.91\% & 66.73\% & 58.19\% & 62.93\%\\
+ \textit{R} + \textit{I}               & 85.63\% & 77.11\% & 81.01\% & 81.80\% & 84.60\% & 91.50\%  &  83.61\%\\  
+ \textit{R} + \textit{I} + \textit{L\&R}         & 86.92\% & 78.13\% & 81.79\% & 82.92\% & 86.42\% & 92.24\%  & 84.74\% \\
\bottomrule
\end{tabular}
\label{tableablation}
\end{table}

By examining the experimental results, we observe that these three components enhance the translation performance of LLMs of code across different parameter scales and translation datasets, highlighting their generalizability. First, The most significant improvement emerges when direct repair is integrated with instrumentation, yielding an absolute improvement in computational accuracy of 12\% to 21\% compared to direct repair without instrumentation across translation datasets. This suggests that for translating lengthy code, instrumentation greatly enhances LLMs of code in capturing and aligning the program states during execution, thus achieving better performance, which aligns with our research motivation. Additionally, localizing inconsistencies in program states and precisely guiding the LLMs of code to re-translate further enhances translation performance, with an improvement of 1\%-4\% in computational accuracy across translation datasets. Although this improvement may seem modest in absolute terms, it is achieved on top of already high performance brought by direct repair and instrumentation. This makes the enhancement particularly noteworthy and underscores the effectiveness of the localize-and-re-translate step.

\begin{boxK}
\small \faIcon{lightbulb} \textbf{Answer to RQ3:} The experimental results reveal that these three components significantly enhance translation performance across code LLMs of varying scales and multiple language pairs, demonstrating broad generalizability. Especially, the results show that integrating instrumentation dramatically improves translation performance, highlighting that capturing and aligning program states plays a vital role in achieving better code translation.
\end{boxK}

\subsection{RQ4: How about LLM translation performance as code length increases?}

To further validate the effectiveness of PAST, we evaluated code LLM performance on different input lengths. The results show the average computation accuracy of code translation across six datasets. These datasets are grouped by source code length into five ranges for evaluation: less than 256 tokens, 256-512 tokens, 512-1024 tokens, 1024-2048 tokens, and more than 2048 tokens.

\Cref{table:model_performance} demonstrates that translation performance for code LLMs of different scales declines as code length increases, likely due to the greater complexity of translating longer code sequences. PAST, however, can partially offset this performance decline, resulting in a more gradual decrease in performance with increasing code length. Moreover, the effectiveness of PAST in mitigating performance loss becomes increasingly pronounced with longer code, as indicated by progressively higher relative improvement rates. Notably, for code lengths exceeding 2048 tokens, PAST achieves a remarkable relative improvement of 97.8\% in computation accuracy for DeepSeekCoder-V2 and 70.20\% for GPT-4o. This highlights PAST’s effectiveness and generalizability in addressing the challenges associated with long code translation.

\begin{table}[h!]
\centering
\footnotesize
\caption{Average computation accuracy of different LLMs across different input sizes.}
\begin{tabular}{cccccc}
\toprule
Model & < 256 & 256 - 512 & 512 - 1024 & 1024 - 2048 & > 2048 \\
\midrule
DeepSeek-Coder-V2 & 69.11\% & 60.91\% & 46.24\% & 33.45\% & 26.83\% \\
+ PAST                & 89.46\% & 84.58\% & 75.13\% & 64.58\% & 53.08\% \\
\rowcolor{gray!20}
\textit{Improvement} & +29.45\% & +38.86\% & +62.48\% & +93.06\% & +97.84\% \\ \midrule
Qwen2.5-72B       & 63.87\% & 50.85\% & 33.28\% & 22.58\% & 18.99\% \\
+ PAST                  & 84.62\% & 72.73\% & 57.73\% & 44.20\% & 33.20\% \\
\rowcolor{gray!20}
 \textit{Improvement}                  & +32.49\% & +43.03\% & +73.47\% & +95.75\% & +74.83\% \\ \midrule
Mistral-Large-2-123B   & 63.98\% & 60.33\% & 46.24\% & 34.00\% & 26.84\% \\
 + PAST                 & 81.39\% & 78.57\% & 71.36\% & 59.51\% & 39.44\% \\
 \rowcolor{gray!20}
  \textit{Improvement}                 & +27.21\% & +30.23\% & +54.33\% & +75.03\% & +46.94\% \\ \midrule
GPT-4o-mini       & 62.25\% & 54.69\% & 44.24\% & 40.58\% & 33.85\% \\
 + PAST                 & 86.99\% & 77.33\% & 66.00\% & 58.51\% & 42.15\% \\
 \rowcolor{gray!20}
  \textit{Improvement}                 & +39.74\% & +41.40\% & +49.19\% & +44.18\% & +24.52\% \\ \midrule
GPT-4o            & 67.75\% & 59.93\% & 52.79\% & 46.26\% & 34.23\% \\
+ PAST                  & 92.43\% & 88.94\% & 81.56\% & 76.55\% & 58.26\% \\
\rowcolor{gray!20}
 \textit{Improvement}                  & +36.43\% & +48.41\% & +54.50\% & +65.48\% & +70.20\% \\
\bottomrule
\end{tabular}
\label{table:model_performance}
\end{table}

\begin{boxK}
\small \faIcon{lightbulb} \textbf{Answer to RQ4:} The experimental results show that PAST significantly improves average computation accuracy across six datasets. While translation performance for code LLMs declines as code length increases, PAST effectively mitigates this drop, demonstrating its effectiveness and generalizability in handling long-code translation challenges.
\end{boxK}

\subsection{RQ5: How about the sensitivity of the approach to the quality of the public tests?}

The quality of public tests can significantly impact the performance of our approach. Public tests with low coverage may fail to reveal potential issues in the Program State Alignment Stage, leading to erroneous programs passing these tests but subsequently failing in the final evaluation based on private tests. To evaluate the sensitivity of our approach to the quality of the public tests, we utilize code LLMs to generate tests to replace the public tests provided in the benchmark. Specifically, we instruct code LLMs to generate five tests per sample based on the original program.  In this context, the code LLMs are primarily tasked with designing the test inputs given the original program.

Comparing the performance of PAST under public tests of varying quality reveals that the tests generated by code LLMs are of insufficient quality. However, as shown in \Cref{lengthexp}, the decline in performance is relatively minor compared to the substantial improvements achieved through our PAST approach. This indicates that while the quality of public tests is important, the robustness of our approach remains largely effective even in less-than-ideal testing conditions.
\vspace{-0.1cm}
\begin{table}[h!]
\centering
\footnotesize
\caption{The sensitivity of our approach to the quality of the public tests.}
\begin{tabular}{lcccccccc}
\toprule
\multirow{2}{*}{Model} & \multirow{2}{*}{Public Tests} & \multicolumn{2}{c}{C++} & \multicolumn{2}{c}{Java} &\multicolumn{2}{c}{Python} & \multirow{2}{*}{Average} \\ \cline{3-4} \cline{5-6} \cline{7-8} 
                   &    & Java   & Python & C++ & Python   & C++ & Java \\ \midrule 

DeepSeekCoder-V2   &   /  & 53.78\% & 48.64\% & 51.28\% & 46.89\% & 54.81\% & 59.31\%  & 52.45\% \\  
+PAST & LLM-generated  & 74.25\% & 72.28\% & 69.10\% & 75.17\% &  65.09\% & 77.94\% & 72.31\%\\  
+PAST & Human-designed & 78.10\% & 74.27\% & 74.12\% & 80.92\% & 71.83\% & 79.33\% & 76.43\% \\  \midrule
Qwen2.5-72B     &  /   & 38.52\% & 39.25\% & 42.33\% & 51.40\% & 50.37\% & 53.69\%  & 45.93\% \\ 
+PAST & LLM-generated    & 63.48\% & 61.64\% & 51.01\% & 68.53\% & 66.30\% & 81.57\% & 65.42\% \\
+PAST & Human-designed & 66.34\% & 63.52\% & 55.17\% & 72.83\% & 69.26\% & 82.84\%  & 68.33\% \\  \midrule
Mistral-Large-2-123B  &   /   & 48.31\% & 46.18\% & 54.47\% & 58.11\% & 50.73\% & 46.25\%  & 50.68\% \\  
+PAST & LLM-generated & 68.25\% & 62.67\% & 61.34\% & 67.97\% & 69.33\% & 76.36\% & 67.65\% \\
+PAST & Human-designed        & 72.13\% & 65.80\% & 65.05\% & 68.42\% & 70.78\% & 76.00\%  & 69.70\% \\  
\bottomrule
\end{tabular}
\label{lengthexp}

\end{table}
\vspace{-0.1cm}

\begin{boxK}
\small \faIcon{lightbulb} \textbf{Answer to RQ5:} The quality of public tests affects translation performance; LLM-generated tests provided weaker feedback in the Program State Alignment Stage, resulting in lower translation accuracy. Nevertheless, PAST still demonstrates a significant improvement in code LLMs across all datasets. This demonstrates the robustness and efficiency of our approach under public tests of varying quality.
\end{boxK}

\section{Discussion}

\subsection{Impact of Data Contamination}

Data contamination arises when instances from a test set are inadvertently included in the training set, compromising the separation between the two datasets. Given that the pre-training and fine-tuning of code LLMs requires a vast amount of code corpus, it is plausible that some translation instances in the test split of our benchmark may have been used during the training of the code LLMs. This overlap could result in an overestimation of the performance of code LLMs.

To address this concern, we conduct an experiment to evaluate the impact of data contamination on the effectiveness of our proposed method. We collect 190 programming problems and 515 solutions from the AtCoder\footnote{https://atcoder.jp/} platform, along with their corresponding test cases from LiveCodeBench~\cite{Jain2024}. Specifically, we gather data from 39 contests held on AtCoder between May 2023 and February 2024. Among the collected solutions, 146 are implemented in C++, 183 in Java, and 186 in Python. 

\begin{figure}[htbp]
    \centering
    \begin{subfigure}[b]{0.48\textwidth}
        \centering
        \includegraphics[width=\textwidth]{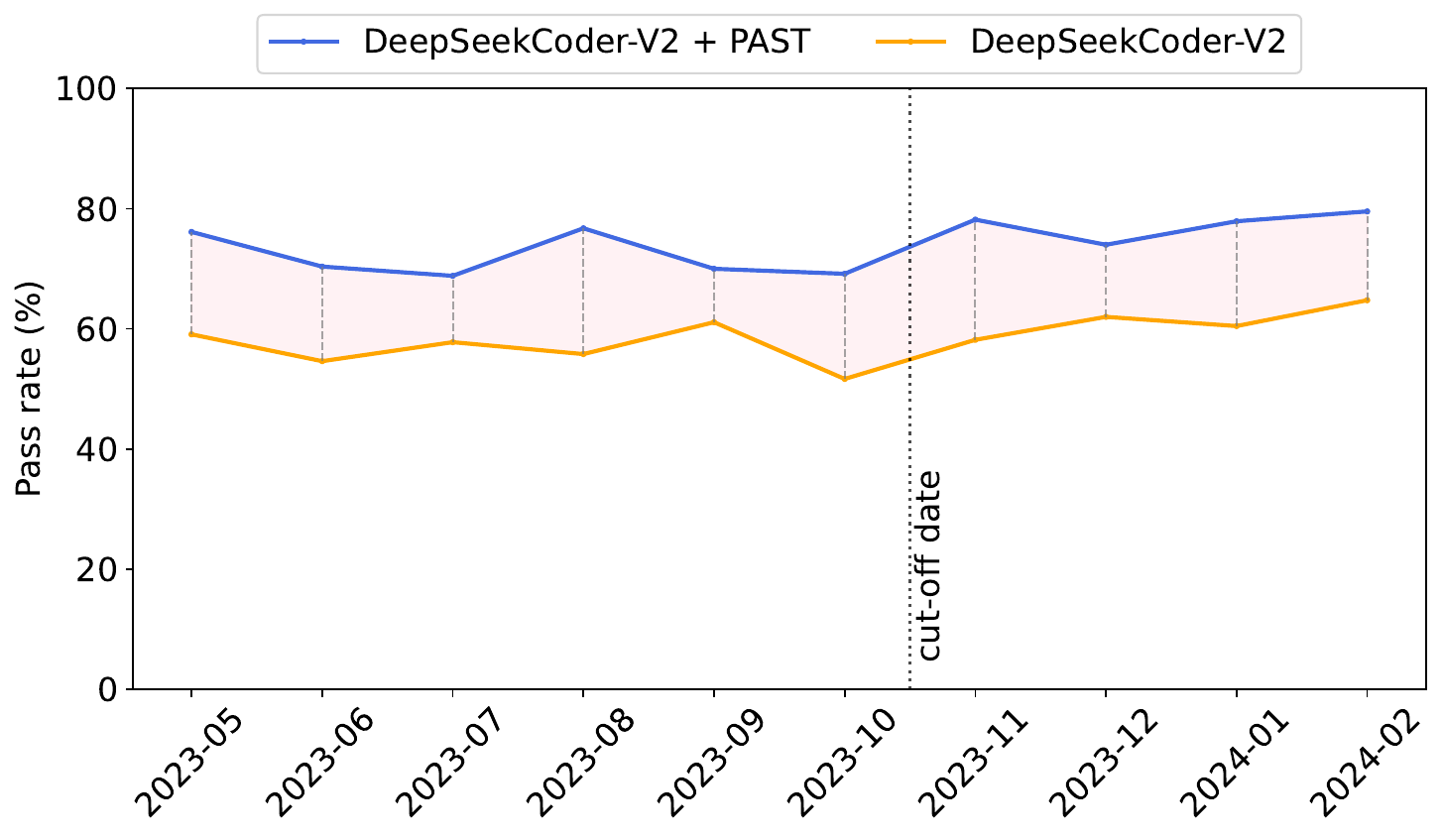}
        \caption{DeepSeekCoder-V2 w. PAST  performs better both before and after the knowledge cut-off date.}
        \label{fig:data_contamination_v2}
    \end{subfigure}
    \hfill
    \begin{subfigure}[b]{0.48\textwidth}
        \centering
        \includegraphics[width=\textwidth]{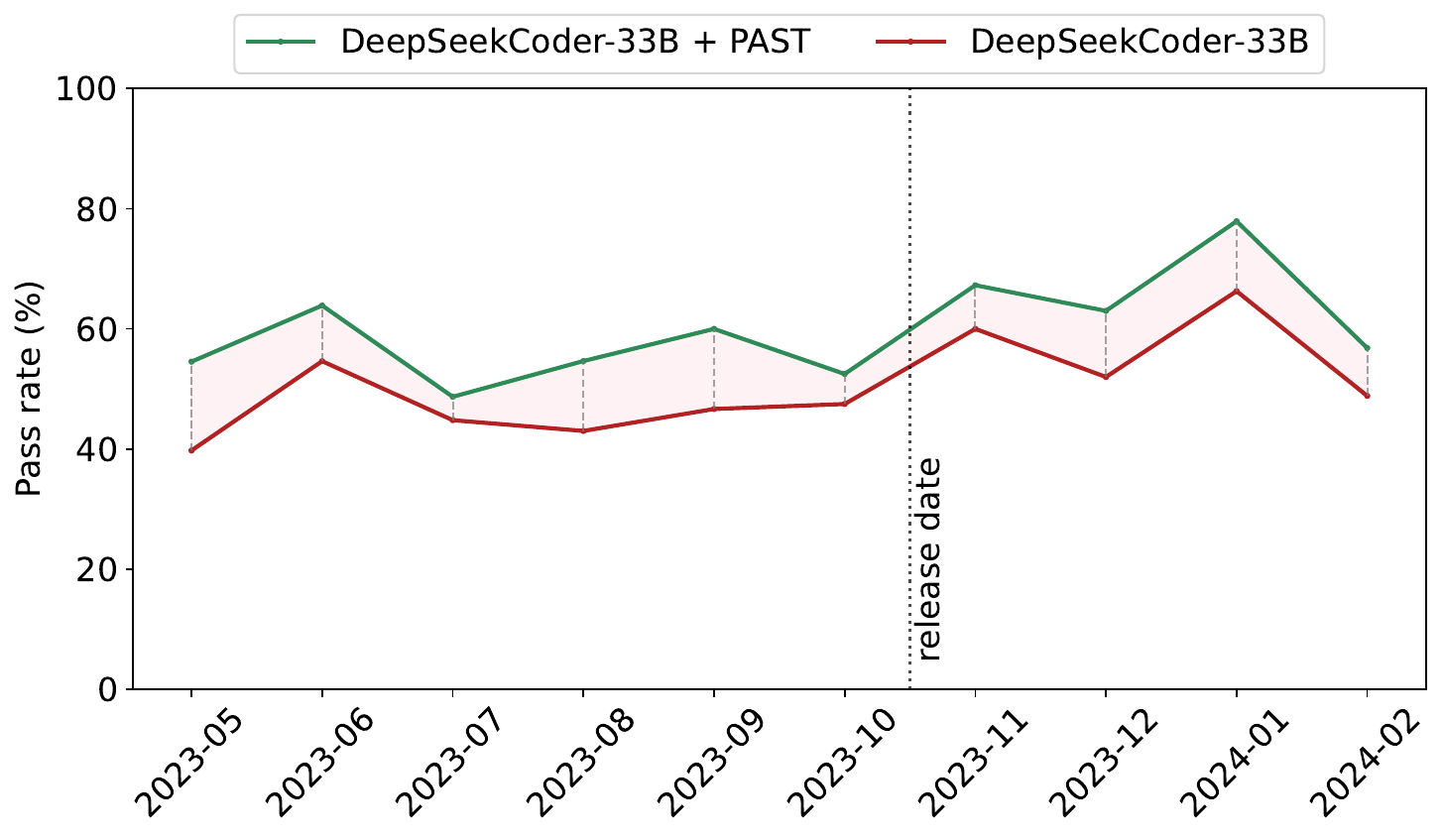}
        \caption{The performance benefit also remains consistent before and after the release date of model weights.}
        \label{fig:data_contamination_33b}
    \end{subfigure}
\end{figure}

For model selection, since the knowledge cut-off dates for Qwen2.5-72B and Mistral-Large-2-123B are not specified in their respective papers, we selected DeepSeekCoder-V2 for our experiment. The knowledge cut-off date for DeepSeekCoder-V2 is November 2023. As shown in \Cref{fig:data_contamination_v2}, the performance benefit of our approach remains consistent both before and after the knowledge cut-off date. To further mitigate the potential impact of data contamination introduced during instruction tuning and other processes after pre-training, which could result in a later knowledge cut-off date, we conducted an additional experiment using DeepSeekCoder-33B, whose model weights were released in November 2023. As demonstrated in \Cref{fig:data_contamination_33b}, the DeepSeekCoder-33B performs consistently better combined with our method. From these two observations, we can confidently conclude that data contamination does not affect the effectiveness of our method.

\subsection{Case Study}

\begin{figure}[htbp]
    \centering
	\includegraphics[width=\textwidth]{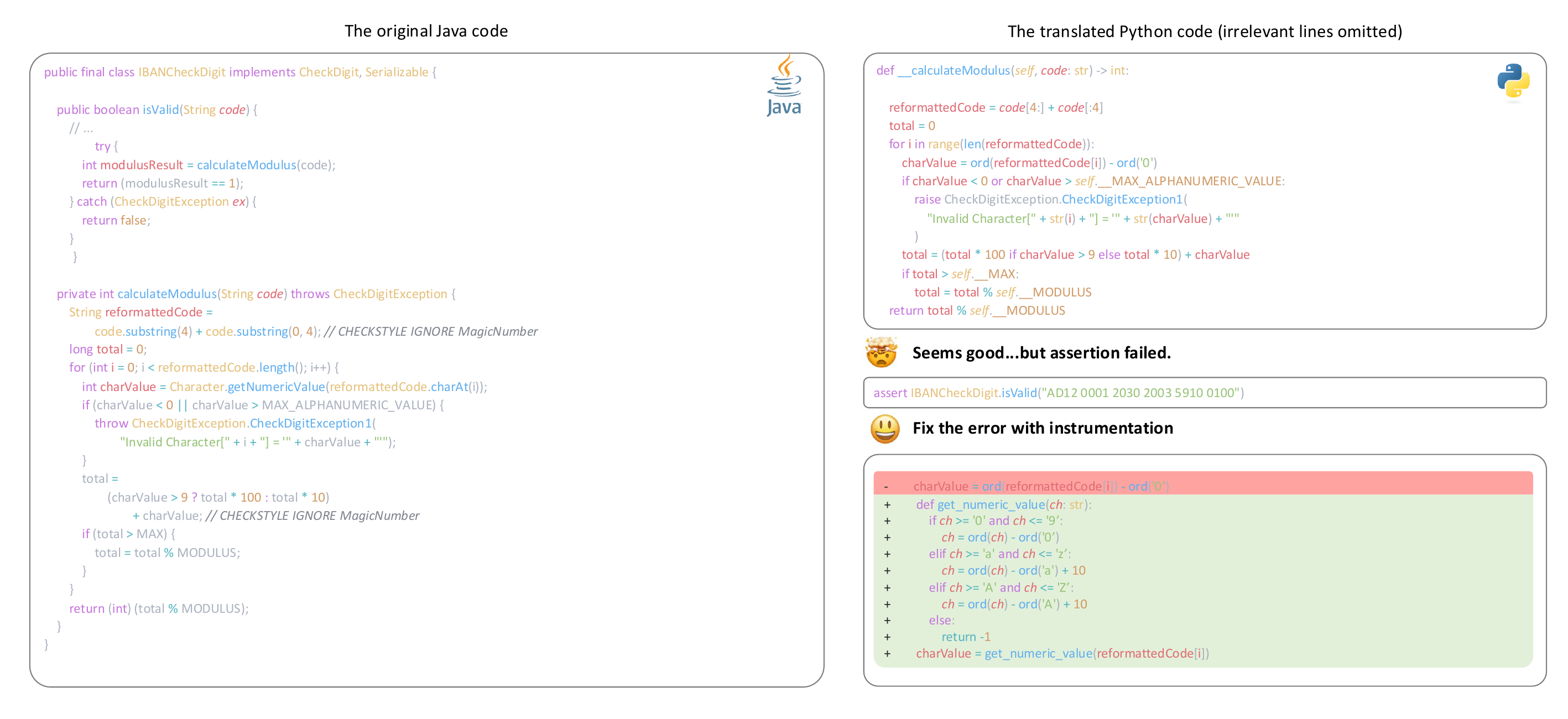}
	\vspace{-2em}
	\caption{The case of our approach in real-world code translation. The Java code is selected from the widely used Apache Commons Validator repository and implements the official IBAN validation standard (ISO 13616). Our approach effectively repairs the error in extracting numeric values from characters in the translated code.}
     \label{fig:case_study}
\end{figure}

To demonstrate our method's ability to solve real-world translation problems, we translate a real-world open-source project, the Apache Commons Validator\footnote{https://github.com/apache/commons-validator}, from Java to Python. For repository-level code translation tasks, directly translating the entire repository can be a long shot and lead to numerous failures. To avoid this, we follow \citet{Ibrahimzada2024} to utilize static analysis to split the repository into code fragments and translate fragments separately. Our approach successfully translates 52 fragments that pass all tests with DeepSeekCoder-V2, outperforming the baseline, which translates only 37 fragments correctly. As shown in \Cref{fig:case_study}, our approach enables the LLM to identify subtle differences between APIs through implementation. While this is still a small step toward fully automated repository translation, it marks significant progress toward that goal.

\section{Threat Analysis}
\label{threat}
Our results are interpreted with three threats to validity in mind.
\begin{itemize}
    \item The internal threat to validity lies in the implementation of compared techniques. To reduce it, we directly reuse the implementation of the compared techniques from their reproducible packages and the weights of pre-trained models, if they are available and executable.
    Otherwise, we reimplement the techniques strictly following the papers on existing mature libraries.
    \item The external threat to validity lies in the dataset used in the experiment. To mitigate the external threat, the widely-used open-source projects are chosen to construct the dataset. Moreover, filtering and preprocessing are performed to ensure no violation case is applicable. In addition, a filtering and preprocessing phase was implemented to eliminate any potential violations or outliers that could skew the results. This process involved rigorously checking for the code solutions and, relevant test cases in the benchmark, thus ensuring that only high-quality data is included for analysis to enhance the reliability of the dataset.
\end{itemize}

\section{Conclusion}

In this paper, we emphasize the insight that the essence of code translation lies in preserving a program's underlying functionality while transforming its appearance from one programming language to another. Nevertheless, existing LLM-based approaches have difficulty in inferring functionality from a program's appearance, resulting in semantic discrepancies between the original and translated program. As demonstrated in our empirical study, this limitation is even more prounouced when faced with complex and lengthy programs. 
To address this issue, we propose a novel approach that leverages intrumentation to explicitly capture and align the program states from entry to exit between the original and translated program. Experimental results show that our approach significantly enhances code LLMs' performance in long code translation, enhancing the usability of code translation in real-world scenarios where program-level and repository-level code migration is more practical. 

This work also empirically investigates code LLM performance in long code translation, identifying a key limitation in existing methods: the inability to fully infer and maintain functionality during translation. Our approach addresses this issue through instrumentation and program state alignment, effectively enhancing translation accuracy for long code. Future work will explore other advanced program analysis techniques, such as symbolic execution, to further refine program state representation. Additionally, integrating our approach into an interactive feedback loop, allowing developers to provide prior knowledge and guide program state alignment, may further boost translation accuracy.

\section{Data Availability}
Our replication package (including code, data, etc.) is publicly available at \url{https://anonymous.4open.science/r/PAST-5411}.

\bibliographystyle{ACM-Reference-Format}
\bibliography{du,li}

\appendix

\end{document}